\documentclass{ws-ijmpa}

\begin{document}

\markboth{A. Rybarska et al.}
{Exclusive photoproduction of $\Upsilon$: from HERA to Tevatron
}

%%%%%%%%%%%%%%%%%%%%% Publisher's Area please ignore %%%%%%%%%%%%%%%
%
\catchline{}{}{}{}{}
%
%%%%%%%%%%%%%%%%%%%%%%%%%%%%%%%%%%%%%%%%%%%%%%%%%%%%%%%%%%%%%%%%%%%%

\title{EXCLUSIVE PHOTOPRODUCTION OF $\Upsilon$: FROM HERA TO TEVATRON
}

\author{ANNA RYBARSKA$^{\star,}$\footnote{E-mail address:
Anna.Cisek@ifj.edu.pl
}
, WOLFGANG SCH\"AFER$^\star$, ANTONI SZCZUREK$^{
\star,\%}$
}
\address{$^\star$ Institute of Nuclear Physics PAN, PL-31-342 Cracow, Poland\\
$^\%$ University of Rzesz\'ow, PL-35-959 Rzesz\'ow,Poland
}

%\author{Wolfgang Sch\"afer}
%\address{Institute of Nuclear Physics PAN, PL-31-342 Cracow, Poland\\
%Wolfgang.Schafer@ifj.edu.pl}

%\author{Antoni Szczurek}
%\address{Institute of Nuclear Physics PAN, PL-31-342 Cracow, Poland\\
%University of Rzesz\'ow, PL-35-959 Rzesz\'ow,Poland\\
%Antoni.Szczurek@ifj.edu.pl}

\maketitle

%\begin{history}
%\received{Day Month Year}
%\revised{Day Month Year}
%\end{history}

\begin{abstract}
The amplitude for photoproduction $\gamma p \to \Upsilon p$
is calculated in a pQCD $k_\perp$-factorization approach.
The total cross section for diffractive $\Upsilon$s is 
compared to recent HERA data.
The amplitude is used to predict the cross section for exclusive
$p \bar p \to p \Upsilon(1S,2S) \bar p$ proces in hadronic reactions at
Tevatron energies. We also included absorption effects.
\end{abstract}

\ccode{PACS numbers: 13.60.Le, 13.85.-t, 12.40.Nn}

\section{Introduction}	

Exclusive production of heavy $Q\bar Q$ vector
quarkonium states in hadronic interactions was never measured, 
but is very attractive from the theoretical
side. Due to the negative charge-parity of the vector meson, 
the Pomeron-Pomeron fusion mechanism of exclusive meson production is not available, 
and instead the production will proceed via photon--Pomeron fusion. 
A possible purely hadronic mechanism would involve the elusive Odderon exchange.
Currently there is no compelling evidence for the Odderon, 
and here we restrict ourselves to the photon-Pomeron fusion mechanism.
The current experimental analyses at the Tevatron
(see, for example, the plenary talk \cite{Pinfold})
call for an evaluation of differential distributions 
including the effects of absorptive corrections.
Predictions for Tevatron require the diffractive amplitude for 
$\gamma p \to \Upsilon p$. This process has been
measured at HERA in the energy range $W \sim$ 100 - 200 GeV \cite{H1}.
This energy range is in fact very much relevant to the exclusive production
at Tevatron energies for not too large rapidities of the meson.

\section{Photoproduction $\gamma p \to \Upsilon p$ at HERA}

The full amplitude for $\gamma p \to \Upsilon p$ process can be written as
(it is explained in ref. \cite{RSS})
\begin{equation}
{\cal M}(W,\Delta^2) = (i + \rho) \, \Im m {\cal M}(W,\Delta^2=0) \, \exp(-B(W) \Delta^2/2) \, ,
\end{equation}
where $\rho$ is a ratio of real and imaginary part of the amplitude. Imaginary part of the 
amplitude depends on the light--cone wave function of $\Upsilon$ 
and the proton's unintegrated gluon distribution (taken from Ivanov-Nikolaev)
\cite{INS06,RSS}.
$B(W)$ is slope parameter which depend on energy :
$B(W) = B_0 + 2 \alpha'_{eff} \log \Big( {W^2 \over W^2_0} \Big) \, $, 
with $\alpha'_{eff} = 0.164$ \ GeV$^{-2}$ , $W_0 = 95$ GeV (see ref. \cite{H1_Jpsi}). 
Our amplitude is normalized to the total cross section:
\begin{equation}
\sigma_{tot}(\gamma p \to \Upsilon p) = {1 + { \rho}^2 \over 16 \pi {B(W)}} \, 
\Big| \Im m { {\cal M}(W,0) \over W^2 } \Big|^2 \, .
\end{equation}
In our calculations we used two types of models for the wave functions:
a Gaussian and a Coulomb--type one, with a power--law tail
in momentum space (ref. \cite{RSS,INS06}). Their parameters were fitted to
the experimental decay widthes $\Upsilon \to e^+ e^-$.
The relevant formalism can be found in refs. \cite{INS06,RSS}. 
It involves the NLO--correction factor $K_{NLO}$.
We have calculated for two different choices of factors $K_{NLO}$. 
In leading order $K_{NLO} = 1$, and next to leading 
order approximation $K_{NLO} = 1 - {16 \over 3 \pi } \alpha_S(m_b^2)$.

\begin{figure}[t]
\begin{center}
\begin{minipage}[t]{0.48\textwidth}
\centerline{\epsfysize 6.4 cm
\epsfbox{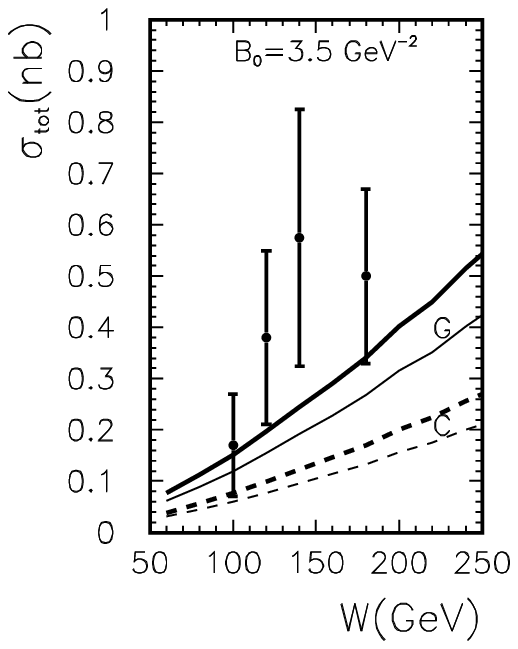}}
\end{minipage} \hfill
\begin{minipage}[t]{0.48\textwidth}
\centerline{\epsfysize 6.4 cm
\epsfbox{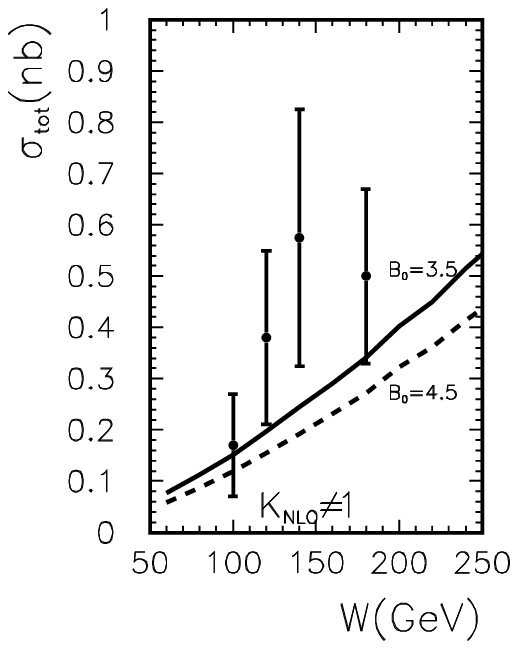}}
\end{minipage}
\vspace{-0.5cm}
\caption{\label{f1} Total cross section for the $\gamma p \to \Upsilon (1S) p$ as a function of
energy. The experimental data are taken from paper
\protect \cite{H1}.
{\bf{Left panel}}: solid curves - Gaussian (G) wave function, dashed curves - Coulomb (C) wave function.
Thick lines were obtained including the NLO correction for the $\Upsilon$ decay width, thin lines are for $K_{NLO}=1$.
{\bf{Right panel}}: solid curves - $B_{0}=3.5 \ GeV^{-2}$, dashed curves - $B_{0}=4.5 \ GeV^{-2}$.}
\end{center}
\end{figure}

In Fig.~\ref{f1} we show the total cross section for the exclusive 
photoproduction $\gamma p \to \Upsilon p$ as a function of the $\gamma-p$
center-of-mass energy $W$.
In the left panel we show results for the two different wave functions:
Gaussian (solid lines) and Coulomb (dashed lines), 
without (thin lines) and with QCD corrections for the
decay width (thick lines).
For $J/\Psi$ photoproduction $B_0$ is $\sim 4.6$ GeV$^{-2}$ (see ref. \cite{SS07}). 
It $B_0$ should be somewhat smaller for the $\Upsilon$ meson.
We show the sensitivity to the slope parameter $B_0$ in the right panel of Fig.~\ref{f1}. 
Our predictions are systematically somewhat below the experimental data. 
The results shown in the right panel of Fig.~\ref{f1} were obtained for the Gaussian wave function 
and include QCD corrections for the decay width.

\section{Exclusive photoproduction in $p \bar p$ collisions}

The full amplitude for $p \bar p$ $\longrightarrow$ $p \bar p$ $\Upsilon$ can be written as
\begin{equation}
\vec {M} ( \vec{p_1}, \vec{p_2} ) = \int{d^2 \vec{k} \over (2 \pi)^2} S_{el}(\vec{k}) \vec{M}^{(0)}
(\vec{p_1} - \vec{k}, \vec{p_2} + \vec{k})
= \vec{M}^{(0)} ( \vec{p_1}, \vec{p_2}) - \delta \vec{M}( \vec{p_1}, \vec{p_2}),
\label{Amplitude}
\end{equation}
where
\begin{equation}
S_{el} \vec{(k)} = (2 \pi)^2 \delta ^{(2)} \vec{(k)} - {1 \over 2} T \vec{(k)},
\ T \vec{(k)} = \sigma ^{p \bar p}_{tot}(s) \exp (-{1 \over 2} B_{el} \vec {k}^2),
\end{equation}
with $B_{el} = 17$ GeV$^{-2}$,\ $\sigma^{pp}_{tot}(s) = 76$ mb 
(see ref. \protect \cite{RSS}).
Here $\vec{p}_1$ and $\vec{p}_2$ are the transverse momenta of outgoing proton and antiproton.

In formula (\ref{Amplitude}) $ \vec{M}^{(0)}(\vec{p_1} , \vec{p_2})$ is 
the Born-amplitude (without absorptive corections) for the process $p \bar p \to p \Upsilon \bar p$ 
which includes our
amplitude for HERA photoproduction and $\delta \vec{M}( \vec{p_1}, \vec{p_2})$ is the absorptive correction. 
Notice, that both proton and antiproton can emit the photon, and these two contributions
interfere in the differential cross section. In particular, the interference is responsible
for a dependence on the azimuthal angle $\phi$ between $\vec{p}_1$ and $\vec{p}_2$.

The differential cross section is given in terms of $\vec{M}$ as
\begin{equation}
d \sigma = { 1 \over 512 \pi^4 s^2 } |\vec{M}|^2\ dy dt_1 dt_2 d \phi,
\end{equation}
where $y$ is the rapidity of the vector meson, $t_{1,2} \simeq -\vec{p}^2_{1,2}$.

\begin{figure}[t]
\begin{center}
\begin{minipage}[t]{0.48\textwidth}
\centerline{\epsfysize 6.4 cm
\epsfbox{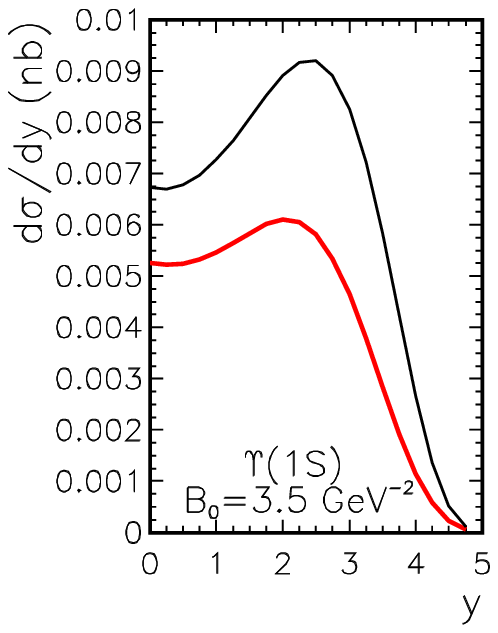}}
\end{minipage} \hfill
\begin{minipage}[t]{0.48\textwidth}
\centerline{\epsfysize 6.4 cm
\epsfbox{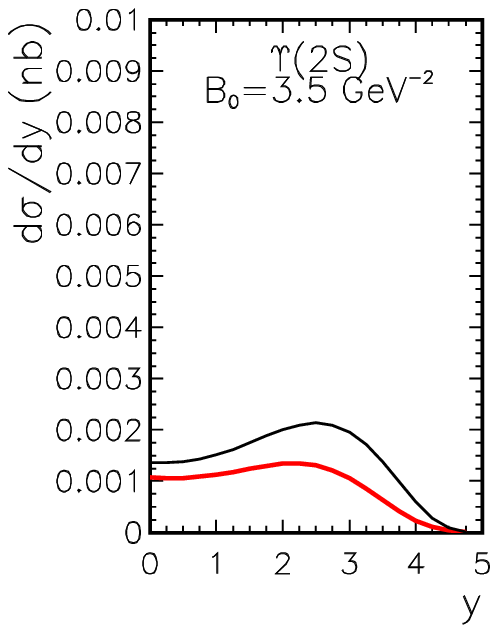}}
\end{minipage}
\vspace{-0.5cm}
\caption{\label{f2} Differential cross section $d\sigma / dy$ for the Tevatron energy
   $W$ = 1960 GeV. {\bf{Left panel}}: results for $\Upsilon (1S)$.
   {\bf{Right panel}}: results for $\Upsilon (2S)$.
  The thin solid line is for the calculation
  with bare amplitude, the thick line for the calculation with
  absorption effects included.}
\end{center}
\end{figure}

The parameters chosen for this calculation correspond to the Gaussian
wave function with $K_{NLO}$ included the QCD corrections.
In Fig.~\ref{f2} we show the distribution in rapidity of $\Upsilon (1S)$
(left panel) and $\Upsilon (2S)$ (right panel). Here the absorption effects cause 
about 20-30\% decrease of the cross section.

\begin{figure}[t]
\begin{center}
\begin{minipage}[t]{0.48\textwidth}
\centerline{\epsfysize 6.4 cm
\epsfbox{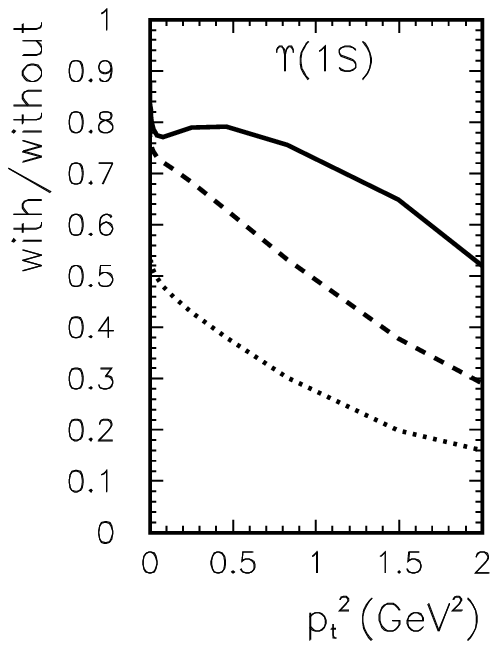}}
\end{minipage} \hfill
\begin{minipage}[t]{0.48\textwidth}
\centerline{\epsfysize 6.4 cm
\epsfbox{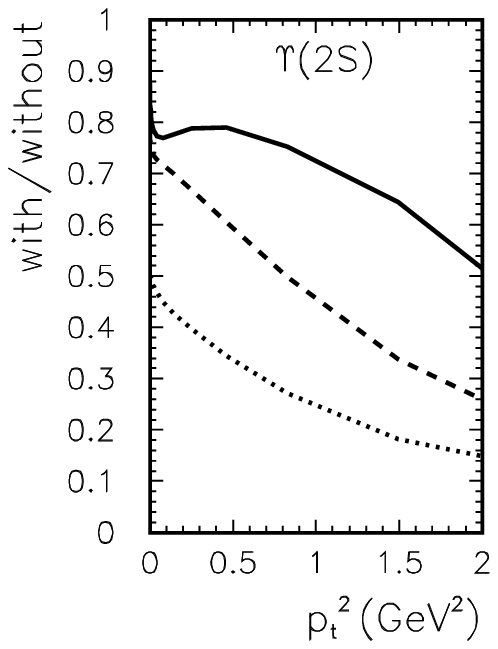}}
\end{minipage}
\vspace{-0.5cm}
\caption{\label{f3} Ratio of $d\sigma/dy dp_t^2$ with absorptive corrections included/
switched off.
{\bf{Left panel}}: Absorption effect for $\Upsilon (1S)$.
{\bf{Right panel}}: the same for $\Upsilon (2S)$. The solid line: $y=0$,
dashed line: $y=2$, dotted line: $y=4$.}
\end{center}
\end{figure}

In Fig.~\ref{f3} we show the ratio of the invariant cross section with to without
absorptive corrections as a function of the $\Upsilon$--transverse 
momentum $p_t$. These results are for different values of rapidity:
$y=0$ (solid lines), $y=2$ (dashed lines) and $y=4$ (dotted lines).  
We can see that absorption effects is bigger for bigger rapidity and also for 
bigger $p_t$.

\section{Conclusions}
The results for $\gamma p \to \Upsilon(1S,2S) p$ production depend on
the model of the wave function. We have compared our results with
a recent HERA data. Our results are somewhat lower than the experimental
data. The amplitudes for the $\gamma p \to \Upsilon p$ process are used
next to calculate the amplitude for the $p \bar p \to p \bar p \Upsilon$
reaction assuming the photon-Pomeron (Pomeron-photon) underlying dynamics.
Absorptive corrections have been included, and they affect the shapes of
various distributions. The resulting cross sections are of measurable size.
%\section*{Acknowledgments}

This work was partially supported by MNiSW under contract 1916/B/H03/2008/34.

%\begin{thebibliography}{000} %for 3 digits
%\begin{thebibliography}{00}  %for 2 digits

\end{document}